\documentclass[12pt]{alt2021}
\usepackage{amssymb,amsmath,amsfonts}
\usepackage{lmodern}

\usepackage{algorithm}
\usepackage{algorithmic}

\usepackage{tikz}

\allowdisplaybreaks

\let\originalleft\left
\let\originalright\right
\renewcommand{\left}{\mathopen{}\mathclose\bgroup\originalleft}
\renewcommand{\right}{\aftergroup\egroup\originalright}

 \newtheorem{thm}{Theorem}
 \newtheorem{cor}{Corollary}
 \newtheorem{prop}{Proposition}

 \newtheorem{problem}{Problem}

\definecolor{pear}{HTML}{c60404}
\definecolor{lightblue}{HTML}{2980B9}

\newcommand{\sset}[1]{\left\{#1\right\}}
\newcommand{\pa}[1]{\left(#1\right)}
\newcommand{\abs}[1]{\left|#1\right|}

\newcommand{\imp}{\Rightarrow}

\newcommand{\CommaBin}{\mathbin{\raisebox{0.5ex}{,}}}
\DeclareMathOperator*{\argmax}{arg\,max}
\DeclareMathOperator*{\argmin}{arg\,min}

\newcommand{\floor}[1]{\left\lfloor#1\right\rfloor}

\newcommand{\R}{\mathbb{R}}

\newcommand{\cF}{\mathcal{F}}
\newcommand{\cG}{\mathcal{G}}

\newcommand{\cO}{\mathcal{O}}
\newcommand{\cP}{\mathcal{P}}

\newcommand{\cX}{\mathcal{X}}

\newcommand{\EE}[1]{\mathbb{E}\left[#1\right]}

\newcommand{\ret}[2]{\pa{#1 \vert #2}}

\newcommand{\eps}{\varepsilon}
\renewcommand{\tilde}{\widetilde}

\makeatletter
\newcommand{\numrel}[2]{
  \refstepcounter{equation}
  \ltx@label{#2}
  \overset{(\theequation)}{#1}
}
\newcommand{\numterm}[1]{\refstepcounter{equation} \ltx@label{#1} (\theequation)}
\makeatother

\newcounter{mylabelcounter}

\makeatletter
\newcommand{\labelText}[2]{%
#1\refstepcounter{mylabelcounter}%
\immediate\write\@auxout{%
  \string\newlabel{#2}{{1}{\thepage}{{\unexpanded{#1}}}{mylabelcounter.\number\value{mylabelcounter}}{}}%
}%
}
\makeatother

\title[Approximation Relationship between RS and DS optimization]{On the Approximation Relationship between Optimizing Ratio of Submodular (RS) and Difference of Submodular (DS) Functions}

\usepackage{times}

\altauthor{%
 \Name{Pierre Perrault} \Email{pierre.perrault@outlook.com}
 \\
 \addr Adobe Research, San Jose, CA --- ENS Paris-Saclay --- Inria Lille%
 \AND
 \Name{Jennifer Healey} \Email{jehealey@adobe.com}
 \\
 \addr Adobe Research, San Jose, CA 
 \AND
  \Name{Zheng Wen} \Email{zhengwen@google.com}
  \\
 \addr DeepMind%
 \AND
 \Name{Michal Valko} \Email{valkom@deepmind.com}
 \\
 \addr DeepMind --- Inria Lille%
}

\begin{document}

\maketitle

\begin{abstract}%
  We demonstrate that from an algorithm guaranteeing an approximation factor for the ratio of submodular (RS) optimization problem, we can build another algorithm having a different kind of approximation guarantee --- weaker than the classical one --- for the difference of submodular (DS) optimization problem, and vice versa. We also illustrate the link between these two problems by analyzing a \textsc{Greedy} algorithm which approximately maximizes objective functions of the form $\Psi(f,g)$, where $f,g$ are two non-negative, monotone, submodular functions and
 $\Psi$ is a {quasiconvex} 2-variables function, which is non decreasing with respect to the first variable. For the choice $\Psi(f,g)\triangleq f/g$, we recover RS, and for the choice $\Psi(f,g)\triangleq f-g$, we recover DS.
 To the best of our knowledge, this greedy approach is new for DS optimization. For RS optimization, it reduces to the standard \textsc{GreedRatio} algorithm that has already been analyzed in \citet{bai2016algorithms}. However, our analysis is novel for this case.
\end{abstract}


\section{Introduction}
Combinatorial optimization is important in many areas of engineering, economics, operations research, computer vision and machine learning. An increasing number of problems coming from this last two areas have been identified as an optimization one involving \emph{submodular} functions (see e.g., \citet{kempe2003maximizing,krause2005near,lin2011-class-submod-sum,bach2011learning}).
Submodular functions can indeed be used in a wide range of applications, such as feature selection, data summarization, or active learning. Learning and maximizing these functions from data is therefore of particular interest \citep{ijcai2018-379,ijcai2019-372}. We can define submodular functions in the following way.
If $[n]\triangleq \sset{1,\dots,n}$ refers a ground set of $n$ elements, and $\cP\pa{[n]}$ is the set of all the subsets of $[n]$, a set function (i.e., a function defined on sets) $f: \cP\pa{[n]} \to \R$ is said to be submodular if for all $A\in \cP\pa{[n]}$ and $B\in \cP\pa{[n]}$, $f(A\cup B) + f(A\cap B) \leq f(A)+f(B)$  \citep{fujishige2005submodular}.  Submodular functions can be characterized by an intuitive \emph{diminishing return} property, where the \emph{marginal gain} $f\ret{i}{A}\triangleq f(A\cup \sset{i}) - f(A)$ of an element $i\in [n]$ in the context of a set $A\in \cP\pa{[n]\backslash\sset{i}}$ is non increasing with respect to $A$. This property is naturally and initially found in economics, where $f(A)$ can measures the
output of a production process involving a set $A$ of different factors.

The class of submodular functions  is special on many aspects. In particular, the problem of minimizing a submodular function is known to be solvable in time polynomial in $n$ \citep{iwata2001combinatorial}, hence submodularity is also often called the discrete analog of convexity \citep{lovasz1983submodular}. In contrast, general submodular maximization is NP-Hard \citep{Schrijver2008}, with many existing algorithms for constant factor approximation (see \citet{buchbinder2017submodular} for a survey).

In addition to problems involving a single submodular function, there exist many problems that are aimed at maximizing (or minimizing) a combination of two submodular functions.
The two functions in this kind of objective generally model two desired behaviours that are more or less complementary to each other. For example, one of the two functions may represent a regularization term on the other (the main objective).
Another common scenario is when one term encodes the gain associated to a particular feasible set $S$, whereas the other represents a cost associated to $S$. 

There are two well known examples of such problems that we are interested in here, involving objectives that is a \emph{difference of submodular} (\nameref{ds}) functions, or a \emph{ratio of submodular} (\nameref{rs}) functions. 
\paragraph{RS and DS optimization problems}
Given two monotone\footnote{A set function $f$ is monotone if $A\subset B \imp f(A)\leq f(B).$}, submodular, but \emph{not necessarily normalized},\footnote{A set function $f$ is said to be normalized if $f(\emptyset)=0$.} functions $f,g :\cP\pa{[n]} \to \R_+$ such that $g$ is positive, we investigate the two following problems involving $f$ and $g$. 
\begin{problem}[\labelText{DS}{ds} Optimization] \citep{kawahara2011prismatic,narasimhan2012submodular,iyer2012algorithms} \[\max_{S\in \cP\pa{[n]}} {f\pa{S}}-{g\pa{S}}.\]\label{pb:diff}
\end{problem}
\vspace{-.4cm}

\begin{problem}[\labelText{RS}{rs} Optimization] \citep{bai2016algorithms,ijcai2017-363,Wang2019}

\[\max_{S\in \cP\pa{[n]}} \frac{f\pa{S}}{g\pa{S}}\cdot\]\label{pb:ratio}

\vspace{-.5cm}
\end{problem}

For DS optimization (but not RS), we can assume that both $f$ and $g$ are normalized without loss of generality. 
We review in the following some recent examples of such optimization problems. We can already see that both problems model the same willingness to maximize the function $f$ while minimizing the function $g$. 
\paragraph{Applications}
RS and DS has been extensively studied in the literature, and can be applied to some real world optimization problems such as sensor placement and feature selection \citep{iyer2012algorithms,bai2016algorithms}. \citet{bai2016algorithms} also highlighted that RS can be used for 
maximizing the F-measure in information retrieval or optimizing
normalized cuts and ratio cuts \citep{shi2000normalized}.
On the other hand, \cite{iyer2012algorithms} considered  modular\footnote{A function $f$ is modular if both $f$ and $-f$ are submodular. This is equivalent to $f(S)=f(\emptyset)+\sum_{i\in S}f\ret{i}{\emptyset}$ for all $S\in \cP([n]).$} upper and lower bounds in order to tackle DS optimization. They also presented some applications such as discriminatively structured graphical models and neural computation or probabilistic inference.
Another application of RS and DS is influence maximization \citep{kempe2003maximizing,chen2010scalable}, where the submodular function $f$ (called \emph{spread} function) is defined on the sets $S$ of vertices of a social network. $f(S)$ represents a gain corresponding to the expected number of users influenced by the seed set $S$ of initial influencers.
Traditionally, the influence maximization problem aims to maximize $f$ under a cardinality constraint on the seed set.
Another formulation is to include a modular function $g$ in the objective of the influence maximization problem, that represents the cost of triggering seed influencers. Notice, both subtracting or dividing $f$ by $g$ have a sens to build the final objective function, that thus falls either in Problem~\ref{pb:diff} or Problem~\ref{pb:ratio}. 

\paragraph{Already known connection between RS and DS}
DS and RS are already known to be connected \citep{bai2016algorithms,chen2019limiting}, since if  $\lambda^*=\max f/g$, then the following are equivalent: \begin{itemize}
    \item $S \in \cP\pa{[n]}$ is a maximizer for DS with the objective function $f-\lambda^* g$,
    \item $S \in \cP\pa{[n]}$ is a maximizer for RS with the objective $f/g$.
\end{itemize}
 We can however give one key difference between the two:   Although
the problem
of finding an optimal solution is NP-hard in both cases \citep{bai2016algorithms, iyer2012algorithms}, RS optimization can be
solved with bounded approximation factors (indeed, \citet{bai2016algorithms} prove an approximation ratio of $\cO\pa{\sqrt{n}\log\pa{n}}$ for the general RS maximization problem, wich is tight up to the $\log$ factor), whereas DS optimization is,
in the worst case, inapproximable \citep{iyer2012algorithms} (actually, they show that any set function maximization problem can be written as a DS optimization problem).
The goal of the present paper is to highlight that both problems are actually \emph{equivalent} in some weaker sense of approximability. 
\paragraph{Contributions}
In Section~\ref{sec:equivalence}, we first prove that the two more general problems of maximizing a ratio $f/g$ and a difference $f-g$ over a general set $\mathcal{X}$ are actually \emph{equivalent} in some sense of approximability, inducing an approximability equivalence result for RS and DS optimization. More precisely, we show that the problem of finding $x\in \mathcal{X}$ having an $\alpha\leq 1$ approximation ratio for $\max f/g$ is equivalent to the problem of finding $x\in \mathcal{X}$ such that $f(x)-g(x)\geq \alpha f(y^*) - g(y^*),$ where $y^*\in \argmax_{y\in \mathcal{X}} {f\pa{y}}-{g\pa{y}}.$ It should be emphasizes here that this last approximation guarantee is weaker than the classical $f(x)-g(x)\geq \alpha \pa{f(y^*) - g(y^*)}$. Such weaker notion of approximation have been considered for instance in \citet{Filmus2012,Sviridenko2013,harshaw2019submodular,perrault2019exploiting,feldman2020guess,ene2020team}.
In the case $g$ has a low \emph{curvature}, we further demonstrate in Section~\ref{sec:greed} the strong relationship between DS and RS by giving approximations guarantees for a simple \textsc{Greedy} algorithm that tackles both problems in a very similar way.
These approximations are new, to the best of our knowledge. This \textsc{Greedy} algorithm was indeed already studied by \cite{bai2016algorithms} for the RS problem, where they claim the same approximation as ours, however, there seems to be an error in their demonstration of Theorem 3.4, since their inequality (16) does not hold for non normalized set functions. In addition, our analysis allows us to extend the result to objectives of the form $\Psi(f,g)$, where 
 $\Psi$ is a {quasiconvex} 2-variables function, which is non decreasing with respect to the first variable.
\section{Approximability of general difference and ratio optimization problems}
\label{sec:equivalence}
The idea behind the connection between RS and DS is actually orthogonal to the fact that the functions $f$ and $g$ are submodular. We thus consider a more general framework here.
 Let $\mathcal{X}$ be a given set, and
let $\mathcal{F}$ be a set of non-negative functions on $\mathcal{X}$, and $\mathcal{G}$ be a cone\footnote{A cone is a subset of a vector space that is closed under multiplication by a positive scalar.} containing positive functions $g$ on $\mathcal{X}$, closed under addition by a scalar $c\in \R$, as long as $g+c$ remains positive. Assume that for $f\in \mathcal{F}$ and $g\in \mathcal{G}$, we can efficiently compute $x_f\in \argmax_\mathcal{X} f$ and $x_g\in \argmin_\mathcal{X} g$. 
Notice, $\mathcal{F}$ can be the set of non-negative monotone submodular functions and $\mathcal{G}$ can be the set of positive monotone submodular functions. 
We consider the two following problems given $\pa{f,g} \in \mathcal{F}\times\mathcal{G}$:
\[(\labelText{\text{Difference}}{diff})\quad \text{Find } x\in \argmax_{x'\in\mathcal{X}} f(x')-g(x').\]
\[(\labelText{\text{Ratio}}{ratio})\quad\text{Find } x\in \argmax_{x'\in\mathcal{X}} f(x')/g(x').\] 
 
Assume an algorithm $\textsc{a}$ that solves exactly \nameref{diff} is available. Then it is already known that one can use a binary search method in order to
 solve \nameref{ratio} with an additive $\varepsilon$ error, using a number of calls to $\textsc{a}$ that is $\cO\pa{\log\pa{1/\varepsilon}}$ \citep{gondran1984graphs}. Indeed, to this aim, we start (using $x_f$ and $x_g$) from an initial interval $[\lambda_-,\lambda_+]$ that contains the optimal ratio $\lambda^*=f(x^*)/g(x^*)$. Then, we take $\lambda=\pa{\lambda_-+\lambda_+}/2$ and call 
$\textsc{a}$ for \nameref{diff} with the objective $f-\lambda g$, to get a solution $x_\lambda$. If $f(x_\lambda)/g(x_\lambda)\leq \lambda$, then by definition of $x_\lambda$, we necessarily have $f(x^*)- \lambda g(x^*) \leq f(x_\lambda)- \lambda g(x_\lambda)\leq 0$ and $\lambda^*\leq \lambda$. Thus,  
$\lambda_+$ is updated as $\lambda_+ \leftarrow \lambda$. Else, by definition of $\lambda^*$, 
$f(x_\lambda)- \lambda g(x_\lambda)\geq 0 \geq f(x_\lambda)- \lambda^* g(x_\lambda)$,  so $\lambda\leq \lambda^*$ and
$\lambda_-$
is updated as $\lambda_- \leftarrow \lambda$. We then go back to the first step with the new shorter interval $[\lambda_-,\lambda_+]$ and iterate this process until the interval is of length less than $\varepsilon$. The last update to $\lambda_-$ corresponds to a solution $x_{\lambda_-}$  which satisfies $f(x_\lambda)/g(x_\lambda)\geq \lambda_-\geq \lambda_+-\varepsilon\geq \lambda^*-\varepsilon. $
The number of iterations needed is clearly of order $\cO\pa{\log\pa{1/\varepsilon}}$, since at each iteration, the size of the interval $[\lambda_-,\lambda_+]$ is divided by $2$.
In summary, we have a fully polynomial-time approximation scheme (FPTAS) algorithm for reducing \nameref{diff} to \nameref{ratio}.

\begin{remark}
In the previous binary search method, ratios $\lambda$ considered are not necessarily of the form $f(x)/g(x)$ for some $x\in \cX$. There exist another iterative method \citep{isbell1956attrition} which works as follows.
Starting from $x_0\in \cX$ and $\lambda=f(x_0)/g(x_0)$, we call the algorithm $\textsc{a}$ to get a maximizer $x_\lambda$ of $f-\lambda g$. If the optimum is lower than $\varepsilon g(x_g)$, then $f(x^*)-(\lambda + \varepsilon) g(x^*)\leq 0$ and $\lambda \geq \lambda^*-\varepsilon$. Otherwise, we set $\lambda \leftarrow f(x_\lambda)/g(x_\lambda)$ and repeat. We obtain an increasing sequence of ratios $(\lambda_k)$ such that $\frac{\lambda_{k+1}-\lambda^*}{\lambda_k-\lambda^*}\leq 1-\frac{g(x^*)}{g(x_{\lambda_{k+1}})} \to_{k\to \infty} 0$
\citep{dinkelbach1967nonlinear,schaible1976fractional}.
\end{remark}

We are not aware of a work that take the reverse path, i.e., that starts from an algorithm $\textsc{a}$ that solves exactly \nameref{ratio} and that uses it for solving \nameref{diff}. The reason is probably because it is usually easier to deal with a difference than a ratio in optimization. However, in the case of RS and DS optimization, it is rather the contrary that happens, since any set function can be written as a difference of two monotone submodular functions \citep{narasimhan2012submodular}, whereas a ratio of monotone submodular functions is a more specific type of set function. In the following, we will provide such reverse path process. We will also state our method for more general approximation versions of the problems \nameref{ratio} and \nameref{diff}, allowing us to make a connection between RS and DS approximability. 

\subsection{FPTAS connection between \nameref{diff} and \nameref{ratio} approximation}
In this subsection, we present a method for designing approximation schemes that make the link between problems \nameref{diff} and \nameref{ratio}. Instead of assuming that an \emph{exact algorithm} is available for one problem, we rather assume that some sort of approximation algorithm is available, which is more general and makes more sense in the context of RS and DS optimization, where existing algorithms are not exact. Specifically, we will provide a FPTAS algorithm for moving from one problem to the other. 
For $\alpha\in [0,1]$, we consider the following approximation problems: Given $\pa{f,g} \in \mathcal{F}\times\mathcal{G}$, find $x\in\mathcal{X}\text{ such that } \forall x'\in\mathcal{X}$:
\[\text{(\labelText{$\alpha$-approximation of Difference}{approx_diff})}\quad f(x)-g(x)\geq \alpha f(x')-g(x').\] 
\[\text{(\labelText{$\alpha$-approximation of Ratio}{approx_ratio})}\quad  f(x)/g(x)\geq \alpha f(x')/g(x').\] 


We state the following Theorem~\ref{thm:eq} which is the main result of this section, and gives a reduction from \nameref{approx_diff} to \nameref{approx_ratio} and vice versa. 

\begin{thm}\label{thm:eq}
Problems \nameref{approx_diff} and \nameref{approx_ratio} are equivalent, in the sense that the first can be reduced with a FPTAS to the second, and conversely. 
\end{thm}

The proof of Theorem~\ref{thm:eq} is postponed to Appendix~\ref{app:thm:eq}.
We use the same dichotomy based FPTAS as we saw previously, for both reductions. For the reduction from  \nameref{approx_ratio} to \nameref{approx_diff}, we use an additive adjustable constant $c$ to $g$ in a similar way as we used the multiplicative constant $\lambda$ previously for the \nameref{approx_diff} to \nameref{approx_ratio} reduction.
The parameter $c$ aims to approach the value $c^*\triangleq\max_{y\in \cX} f(y)-g(y)=f(y^*)-g(y^*)$. Indeed, if $x$ is a maximizer of $f/(g+c^*)$, then 
$f(x)/(g(x)+c^*)\geq f(y^*)/(g(y^*)+c^*) = 1$, so
 $f(x)-g(x)\geq c^*$ and $x$ is a maximizer of $f-g$ as well.

Since we consider approximation algorithms, we will not maintain an interval containing the optimal parameter $\lambda^*=\max_{x\in \cX}f(x)/g(x)=f(x^*)/g(x^*)$ (resp. $c^*=\max_{y\in \cX} f(y)-g(y)=f(y^*)-g(y^*)$), but rather a lower bound on the optimal parameter, and an upper bound on the approximated parameter $\alpha f(x^*)/g(x^*) $ (resp. $\alpha f(y^*) -g(y^*)$).

From Theorem~\ref{thm:eq}, we now clearly have an equivalence between RS and DS in terms of approximability, taking the choices for $\cF$ and $\cG$ as indicated at the beginning of this section. In the following, we refer to the above new concept of approximability for DS optimization as \emph{weak DS approximability}.
We can give a DS example where \nameref{approx_diff} can benefit from an approximation factor from the problem \nameref{approx_ratio}. Indeed, \citet{bai2016algorithms}
has provided an algorithm for \nameref{approx_ratio} with an approximation ratio of $\alpha=\cO(n^{-1/2}\log(n)^{-1})$, using the $\cO(n^{-1/2}\log(n)^{-1})$-approximation of any monotone submodular function in polynomial time by a surrogate function that is the square root of a modular function \citep{goemans2009approximating}. We thus have the following result as a corollary of Theorem~\ref{thm:eq}, giving a solution to the problem \nameref{approx_diff}.

\begin{cor}
Given two monotone, submodular, functions $f,g :\cP\pa{[n]} \to \R_+$, we have an algorithm outputting $S$ with the following  weak DS approximability guarantee: For any $S^*\in \cP\pa{[n]}$, we have
\[f(S)-g(S)\geq \cO\pa{{n^{-1/2}}{\log^{-1}(n)}}f(S^*)-g(S^*).\]
\end{cor}

In the following section, we further investigate approximation for RS and DS in the case where $g$ as a low curvature, i.e., when $g$ is ``close" to be a modular function.

\section{RS and DS problems through \textsc{Greedy} algorithms}\label{sec:greed}
We further illustrate the link between RS and DS problems by proving that a simple \textsc{Greedy} strategy can tackle these two problems in a same way under low curvature assumption on $g$.
We recall the notion of \emph{curvature} \citep{Conforti1984,Vondrak2010,iyer2013curvature,Sviridenko2013} of a monotone submodular function $f$:
 \[c_f\triangleq 1-\min_{i\in [n]}\frac{f\ret{i}{[n]\backslash\sset{i}}}{f\ret{i}{\emptyset}}\in[0,1].\]
Intuitively, the curvature measures how close to modular a submodular set function is. In particular, if $f$ is monotone submodular, then $c_f=0$ if and only if $f$ is modular. We will not state any assumption on the curvature but make the approximation ratio we consider depends on this quantity.

We provide our method in Algorithm~\ref{algo:greedy}. Depending on the function $\Psi$ taken in input, we can deal with a more general class of optimization problems with submodular functions, where RS corresponds to $\Psi(f,g)=f/g$ and DS to $\Psi(f,g)=f-g$.
Remark that before the last step, Algorithm~\ref{algo:greedy} is exactly the standard \textsc{GreedRatio} \citep{bai2016algorithms}. Perhaps surprisingly, we will see that this algorithm can provide guarantees for many kind of ``last step'' without changing the loop part.

We begin by a very structured case where both RS and DS prolems are ``easy''.
We show  in the following Proposition~\ref{prop:mod} that Algorithm~\ref{algo:greedy} is \emph{exact} when functions $f$ and $g$ are modular (but not necessarily monotone), and $\Psi(f,g)=f-g$ or $\Psi(f,g)=f/g$.

\begin{algorithm}[t]
\begin{algorithmic}
\STATE \textbf{Input}: set functions $f,g$, $\Psi.$
\STATE $S_0\leftarrow\emptyset$.
\FOR{$k\in [n]$}
\STATE $i_k\leftarrow\argmax_{i\in [n]\backslash S_{k-1}} {f\ret{i}{S_{k-1}}}/{g\ret{i}{S_{k-1}}}.$
\STATE $S_k\leftarrow S_{k-1}\cup\sset{i_k}.$
\ENDFOR
\STATE $S\leftarrow \argmax_{k\in \sset{0,\dots,n}} \Psi\pa{f\pa{S_k},g\pa{S_k}}.$
\STATE \textbf{Output}: $S.$
\end{algorithmic}
\caption{$\Psi$-\textsc{Greedy}}\label{algo:greedy}
\end{algorithm}

\begin{prop}\label{prop:mod}
If $f$ and $g$ are modular, then Algorithm~\ref{algo:greedy} is exact for $\Psi(f,g)=f-g$ or $\Psi(f,g)=f/g$.
\end{prop}

Proposition~\ref{prop:mod} is proved in Appendix~\ref{app:prop:mod}. We now provide in Theorem~\ref{thm:greed} an approximation guarantee for Algorithm~\ref{algo:greedy} in the general case where $f$ and $g$ are non-negative, monotone and submodular, with a general choice of $\Psi$ that covers both RS and DS. The proof of Theorem~\ref{thm:greed} is delayed to Appendix~\ref{app:thm:greed}.

\begin{thm}\label{thm:greed} Assume $f$ and $g$ are two non-negative, monotone, submodular functions, and assume $\Psi$ is a quasiconvex 2-variables function that is non-decreasing with respect to the first variable. Let $S^*$ be any subset of $[n]$. 
Algorithm~\ref{algo:greedy} is guaranteed to obtain a solution $S$ such that:
 \[\Psi\pa{\pa{1-e^{c_g-1}}f\pa{S^*},g\pa{S^*}} \leq \Psi\pa{f\pa{S},g\pa{S}}.\]
\end{thm}

Notice that $\Psi(f,g)=f-g$ and $\Psi(f,g)=f/g$ both satisfies the assumptions of Theorem~\ref{thm:greed}, i.e., we get the following approximations:
\begin{itemize}
    \item If $\Psi(f,g)=f-g$, \begin{align}\label{rel:DS_cts_approx}\pa{1-e^{c_g-1}}f\pa{S^*} -g\pa{S^*} \leq f\pa{S} -g\pa{S}.\end{align}
    \item If $\Psi(f,g)=f/g$, \begin{align}\label{rel:RS_cts_approx}\pa{1-e^{c_g-1}}\frac{f\pa{S^*}}{g\pa{S^*}}\leq \frac{f\pa{S}}{g\pa{S}}\cdot\end{align}
\end{itemize}
Another example is the objective $\Psi(f,g)=f-\sqrt{g}$, with $g$ modular. A priori, since $\sqrt{g}$ is submodular, we would like to treat this maximization as a simple case of DS. However, $c_{\sqrt{g}}$ might not be small. For example, if $g(S)=\abs{S}$, $c_{\sqrt{g}}=1+\sqrt{n-1}-\sqrt{n}\to_{n\to \infty} 1$. It is thus better to use $\Psi(f,g)=f-\sqrt{g}$, since the approximation from Theorem~\ref{thm:greed} benefits from $c_g=0$ to get $(1-e^{-1})f(S^*)-\sqrt{g(S^*)}\leq f(S)-\sqrt{g(S)}$. 
We get the same improvment with $\Psi(f,g)=f/\sqrt{g}$.
This example can be extended trivially (in both RS and DS) replacing the square root by any non-decreasing concave function.

\paragraph{RS and DS under a knapsack constraint}
We provide here an approximation guarantee similar to that of the Theorem~\ref{thm:greed}, but under a knapsack constraint $g\pa{S}\leq B$, for some budget $B\geq 0$. The approach is slightly different in this case, since it relies on a partial enumeration step for the initialization \citep{khuller1999budgeted}. We provide in Algorithm~\ref{algo:knapgreedy} the modified algorithm. In addition, notice that we also slightly changed the assumption on $\Psi$. We get the following theorem, proved in Appendix~\ref{app:thm:knap}.

\begin{algorithm}[t]
\begin{algorithmic}
\STATE \textbf{Input}: set functions $f,g$, $\Psi,B,\varepsilon.$
\STATE $S_1\leftarrow\argmax_{\abs{S}\leq 2,~g(S)\leq B} \Psi\pa{f(S),g(S)}$.
\STATE $S_2\leftarrow \emptyset$
\FOR{$b\in \sset{1,\dots,\floor{B/\varepsilon}, B/\varepsilon}$}
\FOR{$\abs{S}=3, g(S)\leq B$}
\STATE $N'\leftarrow N\backslash S$
\STATE $S_G \leftarrow$ $\Psi$-Greedy on $N'$ and initialization with $S$, with the condition that $g(S_G)\leq b\varepsilon$.  
\IF{$\Psi(f(S_{G}),g(S_{G}))\geq \Psi(f(S_{2}),g(S_{2})$}
\STATE $S_2\leftarrow S_G$ 
\ENDIF
\ENDFOR
\ENDFOR
\STATE $S\leftarrow \argmax_{k\in \sset{1,2}} \Psi\pa{f\pa{S_k},g\pa{S_k}}.$
\STATE \textbf{Output}: $S.$
\end{algorithmic}
\caption{\textsc{Knapsack} $\Psi$-\textsc{Greedy}}\label{algo:knapgreedy}
\end{algorithm}

\begin{thm}\label{thm:knap} 
Assume $f$ and $g$ are two non-negative, monotone, submodular functions, and
assume that $\Psi$ is non-decreasing with respect to the first variable, and non-increasing with respect to the second variable. Then,
the output $S$ of Algorithm~\ref{algo:knapgreedy} satisfies
  \[\forall S^*\in \cP([n]) ~s.t.~g(S^*)\leq B,~\Psi\pa{\pa{1-e^{c_g-1}}f\pa{S^*},g(S^*)}\leq \Psi\pa{f\pa{S} ,g(S)-\varepsilon}.\]
\end{thm}

\section{Experiments}

In this section, we experimentally evaluate  our Algorithm~\ref{algo:greedy}  for DS maximization, i.e., with $\Psi(f,g)=f-g$. Indeed, an empirical evaluation for $\Psi(f,g)=f/g$ is already given in \citet{bai2016algorithms} (the algorithm is called \textsc{GreedRatio}), and its superiority over other methods such as the Majorization-Minimization and Ellipsoidal Approximation algorithms \citep{bai2016algorithms} has been noticed. 

We compare Algorithm~\ref{algo:greedy} to the state-of-the-art\footnote{The authors noticed that ModMod is fast at each iteration and experimentally does about as well as the Supermodular-Submodular (SupSub) and the Submodular-Supermodular (SubSup) procedures \citep{iyer2012algorithms}.} Modular-Modular (ModMod) procedure of \citet{iyer2012algorithms} on the problem of influence maximization with costly influencers (IMC). In IMC, we consider a social network where some influence interactions between users occur. We record these interactions with a sequence of directed graphs
 $G_t=(V,E_t)$, $t\in [T]$, where
 $e=(u,v)\in E_t$ means that user $u$ \emph{influences} user $v$ on the instance $t$, in other words that some behavior of $u$ (e.g., retweeting or buying a certain product) induced the same behavior for $v$ on the $t$ instance. For a subset of nodes $S\subset V$, we let $f(S)\triangleq \frac{1}{T}\sum_{t=1}^T F(S,G_t),$
 where $F(S,G_t)$ is the number of nodes $u\in V$ such that there is a forward path  in $G_t$ starting from some node of $S$ and ending  to $u$ (i.e., $F(S,G_t)$ is the number of nodes that are \emph{influenced} by $S$).
 It is well known that $S\mapsto F(S,G_t)$ is submodular \citep{kempe2015maximizing}, so is $f$. The function $f$ evaluates how good it is to initially influence the set of node $S$ in order to let the influence spread throughout the network.
We also define the modular function $g(S)=\lambda\sum_{u\in S}c_u,$
where values $c_u$ can be interpreted as a cost to pay to initially influence~$u$. The goal is to maximize the resulting revenue $f-g$.

We consider a  subgraph of Facebook network \citep{snapnets}, with $n=\abs{V}=333$ and $\abs{E}=5038$, as in \citet{wen2017online}, and generate $G_t$ using an independent cascade model \citep{kempe2003maximizing} with weights $w_e$ taken uniformly in in $[0,0.1]$, i.e., $(w_e)_{e\in E}\sim U\pa{0,0.1}^{\otimes E}$. We use $T=10$ in our experiments, and $(c_u)_{u\in V}\sim U\pa{0,1}^{\otimes V}$. For each value of $\lambda$ considered, we run $50$ independent simulations with costs and weights randomly chosen as just described.

Before presenting the result of the experiment, we give more details about the two algorithms we are comparing. We implement a \emph{lazy evaluations} \citep{minoux1978accelerated} version of our greedy algorithm, where instead of maximizing the marginal gain $(f(S\cup\sset{u})-f(S))/c_u$, we
keep an upper bound $\rho(u)$ (initially $\infty$)  on it, sorted in decreasing order.  In each iteration $i$,  we  evaluate  the  element on  top  of  the  list,  say $u$,  and  updates  its  upper  bound with the marginal gain at $S_{i-1}$.   If after the update we have $\rho(u)\geq \rho(v)$ for all $u\neq v$, submodularity guarantees that $u$ is the element with the largest marginal gain. 
The ModMod Algorithm is described as follows. We start with $S_0=\emptyset$, and at each iteration $i\geq 1$, $S_i$ maximizes the modular function $A\mapsto m_{S_{i-1}}(A)-g\pa{A}$, with $A\mapsto m_{X}(A)$ the modular lower bound on $f$ defined by
\[
 m_{X}(\sigma(i)) \triangleq \left\{
    \begin{array}{ll}
        f(\sset{\sigma(1)}) & \mbox{if } i=1 \\
        f\ret{\sigma(i)}{\sset{\sigma(1),\dots,\sigma(i-1)}}& \mbox{otherwise,}
    \end{array}
\right.
\]
where $\sigma$ is a random permutation of $[n]$ such that $\sset{\sigma\pa{1},\dots,\sigma(\abs{X})}=X$. We stop the procedure when $S_{i-1}=S_i$.

\begin{figure}[t]
\centering
\resizebox{\textwidth}{!}{\includegraphics{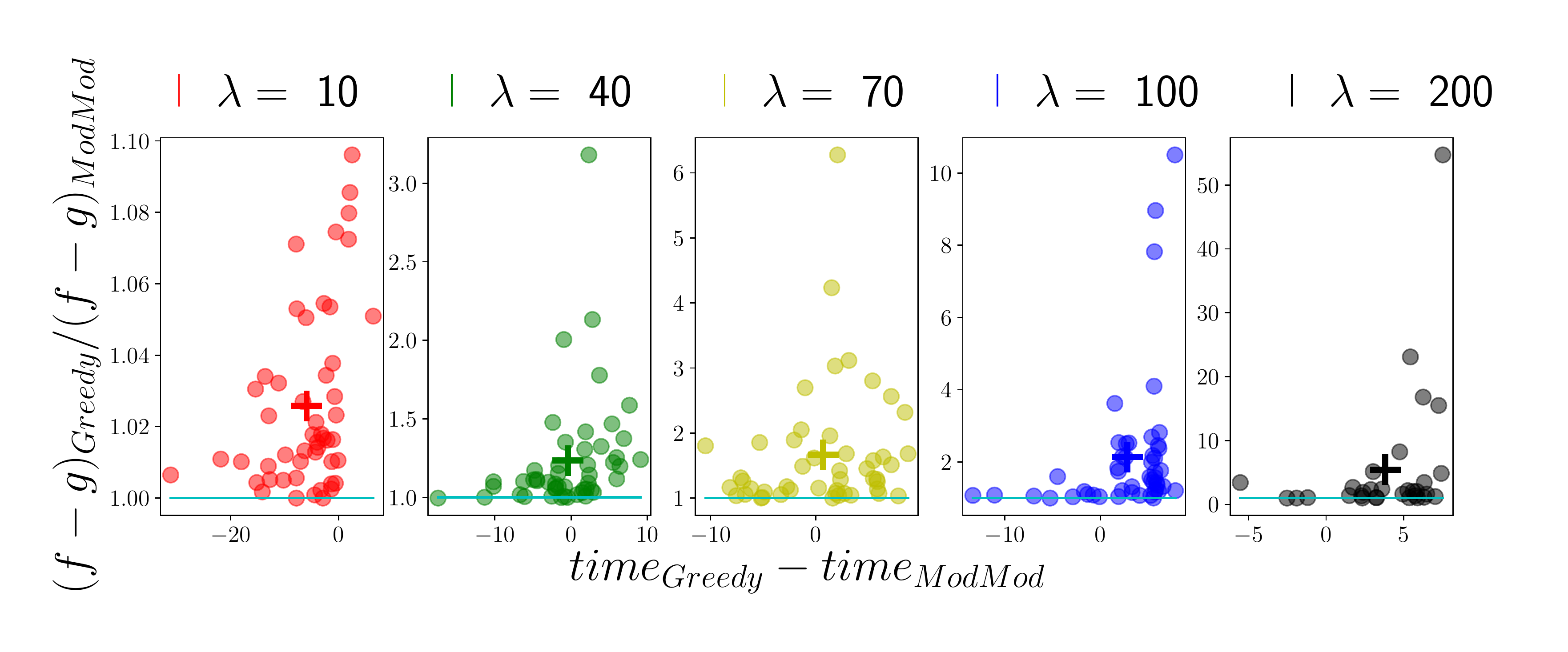}}

\caption{ Performance comparisons between Algorithm~\ref{algo:greedy} and ModMod.
the "+" corresponds to the mean point. The computation time difference is given in seconds.}
\label{exp:rsds}
\end{figure}

Experiments are presented in Figure~\ref{exp:rsds}. We first observe that our Greedy algorithm (Algorithm~\ref{algo:greedy}) is \emph{always} better than ModMod.
More precisely, we see that the larger the value of $\lambda$, the greater the ratio between the two algorithms.
On the other hand, the computation time of the two algorithms seems to be similar, with a slight tendency for the ModMod to be faster when $\lambda$ is bigger (but the performance is worse).

\section{Conclusion}

In this paper, we studied the problem of maximizing the ratio $f/g$ (RS) and the difference $f-g$ (DS), where $f$ and $g$ are two monotonous submodular set functions, as well as the relationship between these two problems.
We showed that for a weaker approximation guarantee, these two problems are equivalent. We also provided a FPTAS to switch from one to the other and generalized the \textsc{GreedRatio} algorithm of \citet{bai2016algorithms} for a larger family of problems including RS and DS. An interesting research direction would be to further generalize the results to combinations of more than two submodular functions. Another interesting question is whether assumptions on $\Psi$ in Theorem~\ref{thm:greed} can be further reduced.

\bibliography{example}
\clearpage
\appendix

\section{Proof of Theorem~\ref{thm:eq}}\label{app:thm:eq}
 
 \begin{proof}
Assume first we have access to an algorithm $\textsc{a}$ for \nameref{approx_ratio}. We consider the procedure given in Algorithm~\ref{algo:diff_ratio}, that take as input $f\in \mathcal{F}$, $g\in \mathcal{G}$ and $\varepsilon>0$.
\newline
\noindent
\begin{minipage}[t]{.5\textwidth}
 \begin{algorithm}[H]
    \caption{\textsc{Difference from Ratio}}\label{algo:diff_ratio}
    $~~~~$\textbf{Initialization}: ${c_-}\leftarrow f(x_g)-g(x_g),$ \\
    $~~~~$ $c_+ \leftarrow\alpha f(x_f)-g(x_g)$,\\
    $~~~~$ $x\leftarrow x_g$.\\
    $~~~~$ \textbf{while} {${c_+ - c_-}>\varepsilon$} \textbf{do}\\
    $~~~~~~~~$ $c\leftarrow\pa{c_+ +c_-}/2$.\\
    $~~~~~~~~$ $y\leftarrow \textsc{a}\pa{f,c+ g} $.\\
    $~~~~~~~~$ \textbf{if} {$f\pa{y}-g\pa{y}\leq c$} \textbf{then}\\
    $~~~~~~~~~~~~$ $c_+ \leftarrow c$.\\
    $~~~~~~~~$ \textbf{else}\\
    $~~~~~~~~~~~~$ $c_- \leftarrow c$.\\
    $~~~~~~~~~~~~$ $x\leftarrow y$.\\
    $~~~~~~~~$ \textbf{endif}\\
    $~~~~$ \textbf{endwhile}\\
    $~~~~$ \textbf{Output}: $x$.
  \end{algorithm}
\end{minipage}%
\begin{minipage}[t]{.5\textwidth}
 \begin{algorithm}[H]
    \caption{\textsc{Ratio from Difference}}\label{algo:ratio_diff}
    $~~~~$\textbf{Initialization}: ${\lambda_-}\leftarrow f(x_g)/g(x_g),$ \\
    $~~~~$ $\lambda_+ \leftarrow\alpha f(x_f)/g(x_g)$,\\
    $~~~~$ $x\leftarrow x_g$.\\
    $~~~~$ \textbf{while} {${\lambda_+ - \lambda_-}>\varepsilon$} \textbf{do}\\
    $~~~~~~~~$ $\lambda\leftarrow\pa{\lambda_+ +\lambda_-}/2$.\\
    $~~~~~~~~$ $y\leftarrow \textsc{a}\pa{f,\lambda g} $.\\
    $~~~~~~~~$ \textbf{if} {$f\pa{y}/g\pa{y}\leq \lambda$} \textbf{then}\\
    $~~~~~~~~~~~~$ $\lambda_+ \leftarrow \lambda$.\\
    $~~~~~~~~$ \textbf{else}\\
    $~~~~~~~~~~~~$ $\lambda_- \leftarrow \lambda$.\\
    $~~~~~~~~~~~~$ $x\leftarrow y$.\\
    $~~~~~~~~$ \textbf{endif}\\
    $~~~~$ \textbf{endwhile}\\
    $~~~~$ \textbf{Output}: $x.$
  \end{algorithm}
\end{minipage}
\newline
\newline

In Algorithm~\ref{algo:diff_ratio}, we always have $c_-\leq f(x)-g(x)$ and $c_+\geq \alpha f(x')-g(x'), ~~\forall x'\in\mathcal{X}$.
    Indeed, these inequalities are clear at the initialization, and the first remains trivially true through the loop from the update of $x$ and $c_-$. The second remains true since for all $x'\in\mathcal{X}$, when $c_+$ is updated, we have  
\[1\numrel{\geq}{rel1}\frac{f(y)}{g(y)+c_+}\numrel{\geq}{rel2} \alpha\frac{f(x')}{g(x')+c_+}\CommaBin\]
where \eqref{rel1} is from $f\pa{y}-g\pa{y}\leq c_+$, and \eqref{rel2} from approximation guarantee of $\textsc{a}$.
Once Algorithm~\ref{algo:diff_ratio} leaves the while loop, $c_+ - c_-\leq\varepsilon$, giving \[\forall x'\in\mathcal{X},\quad \alpha f(x')-g(x')\leq c_+\leq c_-+ \varepsilon\leq f(x)-g(x) + \varepsilon.\] 
Assume now that the algorithm $\textsc{a}$ solves \nameref{approx_diff}, and consider Algorithm~\ref{algo:ratio_diff} that takes as input $f\in \mathcal{F}$, $g\in \mathcal{G}$ and $\varepsilon>0$.

In the same way as previously, in Algorithm~\ref{algo:ratio_diff}, we always have $\lambda_-\leq f(x)/g(x)$ and $\lambda_+\geq \alpha f(x')/g(x'), ~~\forall x'\in\mathcal{X}$, due to the following
   when $\lambda_+$ is updated:  
\[0\numrel{\geq}{rel3}{f(y)}-\lambda_+{g(y)}\numrel{\geq}{rel4} \alpha{f(x')}-\lambda_+{g(x')},\]where \eqref{rel3} is from ${f\pa{y}}/{g\pa{y}}\leq \lambda_+$, and \eqref{rel4} from approximation guarantee of $\textsc{a}$.
Again, once Algorithm~\ref{algo:ratio_diff} leaves the while loop, $\lambda_+ - \lambda_-\leq\varepsilon$, giving \[\forall x'\in\mathcal{X},\quad \alpha f(x')/g(x')\leq \lambda_+\leq \lambda_-+ \varepsilon\leq f(x)/g(x) + \varepsilon.\] 
The number of iterations needed for both Algorithm~\ref{algo:diff_ratio} and \ref{algo:ratio_diff} to converge is of order $\cO\pa{\log\pa{1/\varepsilon}}$ since the quantity $c_+-c_-$ (resp. $\lambda_+-\lambda_-$) is divided by two at each step.
\end{proof}

\begin{remark}
If one want to have an approximation factor in front of $g$ rather than $f$, an idea would be to consider Algorithm~\ref{algo:diff_ratio} with a call of $\textsc{a}(f-c,g)$. However, the function $f-c$ would not be guaranteed to be non-negative, whereas we know in our case that $c+g>0$. Indeed, since $c_-<c_+$ in the while loop, necessarily $c>c_- \geq f(x_g)-g(x_g)$, so $c+g> f(x_g)+g-g(x_g)\geq 0$.  \end{remark}

 \section{Proof of Proposition~\ref{prop:mod}}\label{app:prop:mod}
 
 \begin{proof}
From modularity, a maximizer of $\Psi(f,g)=f-g$ is clearly \[S^*=\sset{i\in [n],~f\ret{i}{\emptyset}-{g\ret{i}{\emptyset}}> 0}=\sset{i\in [n],~\frac{f\ret{i}{\emptyset}}{g\ret{i}{\emptyset}}> 1}.\] 
Again by modularity, ${f\ret{i}{\emptyset}}/{g\ret{i}{\emptyset}}={f\ret{i}{S_{k-1}}}/{g\ret{i}{S_{k-1}}}$ for any $k\in [n]$. So \[S^*=\sset{i_k, k\in [n], \frac{f\ret{i_k}{S_{k-1}}}{g\ret{i_k}{S_{k-1}}} > 1},\] which is of the form $\sset{i_1,\dots,i_k}$ for some $k\in \sset{0,\dots,n}$, so is inspected at the end of Algorithm~\ref{algo:greedy}, which thus outputs a maximizer of $\Psi(f,g)=f-g$.

Let's see now the case $\Psi(f,g)=f/g$. Let $\lambda\triangleq\max_{S\in \cP\pa{[n]}}{f\pa{S}}/{g\pa{S}}$. By the same argument as above, a maximizer of $f-\lambda g$ is
\[S^*=\sset{i_k, k\in [n], \frac{f\ret{i_k}{S_{k-1}}}{g\ret{i_k}{S_{k-1}}} > \lambda},\]
which is of the form $\sset{i_1,\dots,i_k}$ for some $k\in \sset{0,\dots,n}$, so will be inspected at the end of Algorithm~\ref{algo:greedy}. Since the maximum of $f-\lambda g$ is $0$, $S^*$ is also a maximizer of $f/g$, so again Algorithm~\ref{algo:greedy} outputs a maximizer of $\Psi(f,g)=f/g$.
\end{proof}

 \section{Proof of Theorem~\ref{thm:greed}}\label{app:thm:greed}
 \begin{proof}
For all $k\in [n]$,
\begin{align*}\nonumber
    f\pa{S^*} - f\pa{S_{k-1}} &\numrel{\leq}{rel:submo} \sum_{i\in S
    ^*\backslash S_{k-1} }f\ret{i}{S_{k-1}} 
    \\&\numrel{\leq}{rel:algo}\nonumber \frac{f\ret{i_k}{S_{k-1}} }{g\ret{i_k}{S_{k-1}}}
    \sum_{i\in S
    ^*\backslash S_{k-1} } g\ret{i}{S_{k-1}} 
  \\&\numrel{\leq}{rel:submob}\nonumber \frac{f\ret{i_k}{S_{k-1}} }{g\ret{i_k}{S_{k-1}}}
    \sum_{i\in S
    ^* } g\ret{i}{\emptyset} 
    \\&\numrel{\leq}{rel:curv}\nonumber \frac{f\ret{i_k}{S_{k-1}} }{g\ret{i_k}{S_{k-1}}}
    \sum_{i\in S
    ^* } \ \frac{g\ret{i}{S^*\backslash\sset{i} }}{1-c_g} 
    \\&\numrel{\leq}{rel:submot} \frac{f\ret{i_k}{S_{k-1}} }{g\ret{i_k}{S_{k-1}}}
     \cdot \frac{g(S^*)-g(\emptyset)}{1-c_g} 
\end{align*}
Where \eqref{rel:submo} is from submodularity, monotonicity of $f$, \eqref{rel:algo} is from Algorithm~\ref{algo:greedy}, \eqref{rel:submob} is from submodularity, monotonicity of $g$, \eqref{rel:curv} from the curvature of $g$, and \eqref{rel:submot} is from submodularity of $g$.  
In other word, we have for all $k\in[n]$ such that  $f\pa{S^*} - f\pa{S_{k-1}}\geq 0$ that
\begin{align}\frac{\pa{1-c_g} g\ret{i_{k}}{S_{k-1}}}{g(S^*)-g(\emptyset)}\leq \frac{  f\ret{i_k}{S_{k-1}} }{f\pa{S^*} - f\pa{S_{k-1}}}
   \cdot \label{last}
\end{align}

By monotonicity of $g$, there must be an index $\ell\in\sset{0,1,\dots,n-1}$ such that $g\pa{S_\ell}\leq g\pa{S^*}\leq g\pa{S_{\ell+1}}$. Let $\alpha\in [0,1]$ be such that \begin{align}g\pa{S^*} = \alpha g\ret{i_{\ell+1}}{S_{\ell}} +  g(S_{\ell}).\label{combconv}\end{align}  
Now, we claim that \(\pa{1-e^{c_g-1}}{f\pa{S^*}}{}\leq {\pa{1-\alpha}f\pa{S_{\ell}} + \alpha f\pa{S_{\ell+1}}}\) holds. To this end, we will look at two cases:
If
$f\pa{S^*} - \pa{1-\alpha}f\pa{S_{\ell}} - \alpha f\pa{S_{\ell+1}}\leq 0$, then we trivially have 
\[\pa{1-e^{c_g-1}}{f\pa{S^*}}{}\leq{f\pa{S^*}}\leq {\pa{1-\alpha}f\pa{S_{\ell}} + \alpha f\pa{S_{\ell+1}}}.\] 
Otherwise,
\begin{align}\nonumber f\pa{S^*} - \pa{1-\alpha}f\pa{S_{\ell}} - \alpha f\pa{S_{\ell+1}}> 0\\\text{and}\quad f\pa{S^*} - f\pa{S_{k}}> 0\quad\text{for all }k \in[\ell].\label{pos}\end{align}
Thus, we can upper bound $\frac{  f\pa{S^*} - (1-\alpha) f\pa{S_{\ell}} -\alpha f\pa{S_{\ell+1}}}{f\pa{S^*}}$ by the following quantity:
\begin{align*}
&
  \frac{  f\pa{S^*} - (1-\alpha) f\pa{S_{\ell}} -\alpha f\pa{S_{\ell+1}} }{f(S^*)-f(\emptyset)} \\&=\frac{  f\pa{S^*} - f\pa{S_{\ell}} -\alpha f\ret{i_{\ell+1}}{S_\ell} }{  f\pa{S^*} - f\pa{S_{\ell}}  }\prod_{k\in[\ell]}\frac{  f\pa{S^*} - f\pa{S_{k}} }{f\pa{S^*}- f\pa{S_{k-1}}} 
  \\&= \pa{1-\frac{  \alpha f\ret{i_{\ell+1}}{S_{\ell}} }{f\pa{S^*} - f\pa{S_{\ell}}}}
  \prod_{k\in[\ell]}\pa{1-\frac{  f\ret{i_k}{S_{k-1}} }{f\pa{S^*} - f\pa{S_{k-1}}}}
  \\&\numrel{\leq}{rel:prev}
  \pa{1-\frac{\alpha\pa{1-c_g} g\ret{i_{\ell+1}}{S_{\ell}}}{g(S^*)-g(\emptyset)}}
  \prod_{k\in[\ell]}\pa{1-\frac{\pa{1-c_g} g\ret{i_{k}}{S_{k-1}}}{g(S^*)-g(\emptyset)}}
  \\&\numrel{\leq}{rel:exp}
  \exp\pa{-\pa{1-c_g}\frac{{ \alpha g\ret{i_{\ell+1}}{S_{\ell}} + \sum_{k\in[\ell]} g\ret{i_{k}}{S_{k-1}}}}{g(S^*)-g(\emptyset)}} 
    \\&= \exp\pa{-\pa{1-c_g}\frac{{ \alpha g\ret{i_{\ell+1}}{S_{\ell}} +  g(S_{\ell})-g(\emptyset)}}{g(S^*)-g(\emptyset)}    }
    \\&
   \numrel{=}{rel:combc}e^{c_g-1}. 
\end{align*}
Where \eqref{rel:prev} is from $\eqref{last}\text{ and }\eqref{pos}$, \eqref{rel:exp} is from $1-x\leq e^{-x}$ and \eqref{rel:combc} is from \eqref{combconv}.
Rearranging the inequality, we obtain our claim:
\begin{align*}
    \pa{1-e^{c_g-1}}f\pa{S^*} \leq  (1-\alpha) f\pa{S_{\ell}} +\alpha f\pa{S_{\ell+1}}.
\end{align*}
We now use the fact that $\Psi$ is non-decreasing with respect to the first variable to get
\[\Psi\pa{\pa{1-e^{c_g-1}}f\pa{S^*},\cdot}\leq \Psi\pa{(1-\alpha) f\pa{S_{\ell}} +\alpha f\pa{S_{\ell+1}},\cdot}.\]
In the previous inequality, we evaluate the second variable to $g(S^*)=\pa{1-\alpha}g\pa{S_{\ell}} + \alpha g\pa{S_{\ell+1}}$, and use the quasiconvexity of $\Psi$ to get that $\Psi\pa{\pa{1-e^{c_g-1}}f\pa{S^*},g(S^*)}$ is upper bounded by
\begin{align*}&\Psi\pa{(1-\alpha) f\pa{S_{\ell}} +\alpha f\pa{S_{\ell+1}},\pa{1-\alpha}g\pa{S_{\ell}} + \alpha g\pa{S_{\ell+1}}}
\\&\leq \max_{k\in \sset{\ell,\ell+1}}\Psi\pa{f\pa{S_{k}},g\pa{S_{k}}}
.\end{align*}
To conclude, see that the output $S$ of Algorithm~\ref{algo:greedy} maximizes $\Psi\pa{f\pa{S_{k}},g\pa{S_{k}}}$ over $k\in \sset{0,\dots,n}$, so we have $\max_{k\in \sset{\ell,\ell+1}}\Psi\pa{f\pa{S_{k}},g\pa{S_{k}}}\leq \Psi\pa{f\pa{S},g\pa{S}}$.
\end{proof}

\section{Proof of Theorem~\ref{thm:knap}}\label{app:thm:knap}

\begin{proof} Without loss of generality, we assume that $\abs{S^*}\geq3$, since otherwise the algorithm find the optimal solution. 
Let us write $S^*_i=\sset{e^*_1,\dots,e^*_i}$ where the elements are ordered such that:
$$e^*_j=\argmax_{e\in S^*\backslash S^*_{i-1}} f(S^*_{i-1}\cup\sset{e}).$$
We consider 
the iteration where $b$ is the largest such that $(b-1)\varepsilon \leq g(S^*)\leq b\varepsilon$
and where $S=S^*_3$.
We thus have
$g(S_G)-\varepsilon\leq (b-1)\varepsilon\leq g(S^*)\leq b\varepsilon\leq g(S_G\cup \sset{x^*})$ for $x^*=\argmax_{i\in S^*\backslash S_{G}} {f\ret{i}{S_{G}}}/{g\ret{i}{S_{G}}}$.
Replacing $i_{\ell+1}$ by $x^*$ in the proof of Theorem~\ref{thm:greed} and letting $\alpha$ being such that $g(S^*)=(1-\alpha)(g(S_G)-\varepsilon)+\alpha g(S_G\cup\sset{x^*})$, we can replace equality \eqref{rel:combc} with an inequality and have that
\[\pa{1-e^{c_g-1}}{\tilde f\pa{S^*}}\leq {\pa{1-\alpha}\tilde f\pa{S_{G}} + \alpha \tilde f\pa{S_{G}\cup \sset{x^*}}}\leq \tilde f\pa{S_{G}\cup \sset{x^*}} ,\]
where $\tilde f(S)\triangleq f(S\cup S^*_3)-f(S^*_3)+f(\emptyset)$.
Notice now that from the definition of $S^*_3$, and the submodularity of $f$, we have $ f\pa{S_{G}\cup\sset{x^*}}-f\pa{S_{G}}\leq \frac{f(S_3^*)-f(\emptyset)}{3} $, we thus have
\[\pa{1-e^{c_g-1}}{f\pa{S^*}}\leq\pa{e^{c_g-1}-1/3}\pa{f(S^*_3)-f(\emptyset)}+\pa{1-e^{c_g-1}}{f\pa{S^*}}\leq f\pa{S_{G}} .\] 
We now use the fact that $\Psi$ is non-decreasing with respect to the first variable, and non-increasing with respect to the second variable
\[\Psi\pa{\pa{1-e^{c_g-1}}f\pa{S^*},g(S^*)}\leq \Psi\pa{f\pa{S_{G}} ,g(S_G)-\varepsilon}.\]
Notice that $S_G$ satisfies the knapsack constraint since $g\pa{S_{G}}\leq b\varepsilon\leq B.$ We thus have the desired result.
\end{proof}

\section{Optimality of the approximation when $g$ is modular}
In this section, we show that the approximations results given in \eqref{rel:DS_cts_approx} and \eqref{rel:RS_cts_approx} are optimal. To this end, we shall prove that if there exists an algorithm with an improved $\alpha=1-1/e+\eps$ ratio, for some $\eps>0$, then we can use it to solve a randomized version of submodular maximization under cardinality constraint, with a factor $\alpha$. We consider the DS problem only, noticing that the same proof holds for RS. Let $f$ be a monotone normalized submodular function. Given some $\lambda>0$, we can use a linear search to find the value $\lambda'=h(\lambda)$ that maximizes $f(S_{\lambda'})-\lambda\abs{S_{\lambda'}}$, where $S_{\lambda'}$ is 
the $\alpha$-approximate maximizer such that $f(S_{\lambda'})-\lambda'\abs{S_{\lambda'}}\geq \alpha f(S^*)-\lambda'\abs{S^{*}}$. We notice that we have 
\[f(S_{h(\lambda)})-\lambda\abs{S_{h(\lambda)}}\geq f(S_{\lambda})-\lambda\abs{S_{\lambda}}\geq \alpha f(S^*)-\lambda\abs{S^{*}}.\]
In addition, we also have that the function $\lambda \mapsto \max_{\lambda'}f(S_{\lambda'})-\lambda\abs{S_{\lambda'}} $ is continuous, and non-increasing. If $\lambda$ is  some  branching  point  between two  maximizers $S_{h(\lambda^+)}$ and $S_{h(\lambda^-)}$, then
looking at the derivative with respect to $\lambda$, we see that the cardinality of the maximizer dictates which one dominates if we increase/decrease $\lambda$. This means that $\lambda\mapsto \abs{S_{h(\lambda)}}$ is non-increasing. Thus, given an integer $m\in [n]$, we have that there is a $\lambda$ (that we can find with a binary search) such that either $\abs{S_{h(\lambda)}}=m$, or $\abs{S_{h(\lambda
^+)}}<m<\abs{S_{h(\lambda
^-)}}$ (in such case, we use a randomized maximizer $\tilde S_{h(\lambda)}$ so that $\EE{\abs{\tilde S_{h(\lambda)}}}=m$). We finally see that
\[ \EE{f\pa{\tilde S_{h(\lambda)}}} \geq \alpha f(S^*), \text{ with }\EE{\abs{\tilde S_{h(\lambda)}}}=m,~\forall S^* \text{ s.t. }\abs{S^*}=m. \]Note that the linear and binary search gives a slight error in the $\alpha$ factor, but that this error can be made strictly smaller than $\eps$, thus contradicting the optimality of the factor $1-1/e$ for the above problem.

\end{document}